----------
X-Sun-Data-Type: default
X-Sun-Data-Description: default
X-Sun-Data-Name: solitonic
X-Sun-Charset: us-ascii
X-Sun-Content-Lines: 1583


\magnification=1200
\settabs 18 \columns

\baselineskip=17 pt

\def\s{\smallskip}

\def\sqr#1#2{{\vcenter{\vbox{\hrule height.#2pt
 \hbox{\vrule width.#2pt height#1pt \kern#1pt
 \vrule width.#2pt} \hrule height.#2pt}}}}
\def\square{\mathchoice \sqr65 \sqr65 \sqr{2.1}3 \sqr{1.5}3}

\def\operp{\hbox{${\kern+.25em{\bigcirc}
\kern-.85em\bot\kern+.85em\kern-.25em}$}}

\def\lsim{\;\raise0.3ex\hbox{$<$\kern-0.75em\raise-1.1ex\hbox{$\sim$}}\;}
\def\gsim{\;\raise0.3ex\hbox{$>$\kern-0.75em\raise-1.1ex\hbox{$\sim$}}\;}
\def\no{\noindent}

\def\ce{\centerline}
\def\ve{\vfill\eject}
\def\rdots{\mathinner{\mkern1mu\raise1pt\vbox{\kern7pt\hbox{.}}\mkern2mu
 \raise4pt\hbox{.}\mkern2mu\raise7pt\hbox{.}\mkern1mu}}

\def\e e{$e^+ e^-$ }


\rightline{UCLA/97/TEP/8}
\rightline{April 1997}
\vskip.5cm

\ce{\bf DYONIC BLACK HOLES AND RELATED SOLITONS}
\vskip.75cm

\ce{A. C. Cadavid}
\s
\ce{\it Department of Physics}
\ce{\it California State University, Northridge, CA 91330}
\vskip.3cm
\ce{R. J. Finkelstein}
\s
\ce{\it Department of Physics and Astronomy}
\ce{\it University of California, Los Angeles, CA 90095-1547}
\vskip1.0cm

\no {\bf Abstract.}  There is a growing literature on 
dyonic black holes as
they appear in string theory.   
Here we examine the correspondence limit
of a dyonic black hole which is not supersymmetric.  Assuming the existence
of a dyon with non-supersymmetric Kerr-Schild structure, we calculate its
gravitational and electromagnetic fields and compute its mass and angular
momentum to obtain a modified B.P.S. relation.  The contribution of the
angular momentum to the mass appears in the condition for the appearance of
a horizon.  One of the advantages of the Kerr-Schild frame is the possibility
of a Lorentz covariant treatment since gravitational pseudo-energy-momentum
tensor vanishes in this frame.  

The solutions coming from string theory exhibit a central singularity.
We briefly discuss the possibility that there are true
solitonic solutions free of all singularities.  We would expect these solitons
to show noding radial behavior in contrast to the known stringy black holes.
\ve

\line{{\bf 1. Introduction.} \hfil}
\s

At the level of classical field theory and special relativity, theoretical
models of the elementary particles have infinite mass unless they are 
solitonic.  However, dyonic solitons do appear naturally in particular 
non-Abelian field theories$^1$ and at the level of special relativity satisfy
the B.P.S. relation$^2$ connecting mass, electric and magnetic charge:
$$
m^2\geq e^2+g^2 \eqno(1.1)
$$
\no There is a similar bound that has been established at the general
relativistic level, namely$^3$
$$
m^2\geq G^{-1}(e^2+g^2)~. \eqno(1.2)
$$
\no The extension to general relativity is obviously necessary since a 
satisfactory description of elementary particles must contain gravitational
couplings, and a natural candidate for an elementary particle is possibly a
solitonic version of a black hole.  In recent work there have been many
attempts to understand these putative particles, including spinning black holes,
as they appear in higher dimensional and locally supersymmetric theories.$^4$
This work has also led to interesting conjectures about the Bekenstein-Hawking
entropy of black holes.

The particle-like solutions of the field equations, the so-called solitons,
coming from string theories appear always to exhibit central singularities.  In
this respect they resemble Schwarzschild black holes and differ from the
original idea of a soliton as a classical lump of field with no singularities.
We are here concerned mainly with the black-hole type of particle but we shall
also briefly consider the possibility of singularity free solitons.

We shall study the rotating dyon at the general relativistic level without
the complications of higher dimensionality and local supersymmetry.  Our
speculative input will be confined 
to the assumptions that dyons$^5$ do exist and
may be described by a Kerr-Schild structure.$^6$  We should also like to compare
the mass of this specific structure with that predicted by the general relations
(1.1) and (1.2).

One of the advantages of the Kerr-Schild representation of a spinning source
is the possibility of a Lorentz covariant treatment$^7$ since the gravitational
pseudo-energy-momentum tensor (p.e.m.t.) vanishes in this representation.$^8$
Passing from a general coordinate system to Kerr-Schild coordinates therefore
cancels the gravitational energy and momentum and may be interpreted as a kind
of acceleration according to the equivalence principle.  Additional
Poincar\'e transformations will not change the Kerr-Schild metric.  There are
also linear but complex translations which lead from the neutral spinning source
to either the Schwarzschild source$^9$ or to the charged spinning source.$^{10}$
Here we shall use the method of complex translation to obtain a description of 
a 4-dimensional dyon.

An important role in the considerations of this paper is played by the
gravitational energy.  Since gravitational energy is not localizable, there is
an arbitrariness in discussing it and consequently there have been many
different proposals for the total energy-momentum 
of an isolated system.$^{11}$  These different
expressions for the pseudo-energy-momentum tensor all lead to energy-momentum
vectors that may be written as esentially equivalent surface integrals.  The problem has been discussed
in generality by Arnowitt, Deser, and Misner.$^{12}$  Our problem is
simpler since we are assuming not only the Kerr-Schild metric but also time
independence.  We shall show that in this metric the contribution of the
gravitational field to the pseudo-energy-momentum tensor vanishes exactly if
the source field is conformal (traceless).

If one takes the view that string theory is essentially correct, the first
part of this paper may be regarded as the correspondence limit of a higher
dimensional construction such as $M$ theory.

The second part of this paper distinguishes between string
 solitons and
``true" solitons as candidates for elementary particles.
\ve

\line{{\bf 2. The General Relativistic Structure of a Rotating Dyon.} \hfil}
\s

Since the dyon is the source of both an electric (e) and magnetic (g) charge,
it is also the source of two independent fields, $F^{(e)}_{\mu\nu}$ and
$F^{(g)}_{\mu\nu}$ with associated vector potentials $A_\mu^{(e)}$ and
$A_\mu^{(g)}$ as well as energy-momentum tensors $\theta_{\mu\nu}^{(e)}$ and
$\theta_{\mu\nu}^{(g)}$.  For $\theta_{\mu\nu}^{(A)}$ we have the usual
construction
$$
\theta_{\mu\nu}^{(A)} = \biggl(F_\mu^{~\sigma}F_{\nu\sigma} -
{1\over 4} g_{\mu\nu} F^{\alpha\beta}F_{\alpha\beta}\biggr)^A \qquad
A = (e,g) \eqno(2.1)
$$
\no and the complete energy momentum tensor of the electromagnetic field is
$$
\theta_{\mu\nu} = \theta_{\mu\nu}^{(e)} + \theta_{\mu\nu}^{(g)}~. \eqno(2.2)
$$
\no We do not assume that these fields are generated by a non-Abelian theory.

We are also assuming that this dyon is the rotating source of a gravitational
field, $g_{\mu\nu}$, which may be written in the Kerr-Schild form:$^6$
$$
g_{\mu\nu} = \eta_{\mu\nu} - 2m\ell_\mu\ell_\nu \eqno(2.3)
$$
\no where $\eta_{\mu\nu}$ is the Minkowski metric (1,-1,-1,-1) amd where the
null vector $\ell_\mu$ is
$$
\eqalignno{&\ell_\mu = (\ell_o,\ell_o\lambda_k) & (2.4a) \cr
&\lambda_i\lambda_i = 1~. & (2.4b) \cr}
$$
\no We shall show that$^7$
$$
\ell_o^2 = \biggl(1 - {e^2+g^2\over 2m\rho}\biggr) \alpha(\rho) \eqno(2.5a)
$$
\no where $\alpha$ is the real part of a harmonic function:
$$
\gamma = \alpha + i\beta \eqno(2.5b)
$$
\no and where
$$
\rho = {\alpha\over\alpha^2 + \beta^2}~. \eqno(2.5c)
$$
\no Thus $\ell_o^2$, and therefore $g_{\mu\nu}$, is entirely fixed by the 
harmonic function $\gamma$.

In the uncharged case $e=g=0$ and
$$
\ell_o^2 = \alpha~. \eqno(2.6)
$$
\no In this case $\ell_o^2$ may be regarded as a generalization of the
Newtonian potential, while $\beta$, the imaginary part of $\gamma$, is
proportional to the specific angular momentum of the source.

Instead of describing $\ell_\mu$ in terms of $\gamma$ we may describe it in
terms of its reciprocal, $\omega$, which may be expressed as a complexified
radial coordinate:
$$
\eqalignno{\omega &= \bigl[x^2+y^2+(z-ia)^2\bigr]^{1/2} & (2.7) \cr
&= \rho + i\sigma ~. & (2.8) \cr}
$$
\no Then $\rho$ may be regarded as a new coordinate substituting for the usual
radial coordinate, $r$, and $\sigma$ as a new coordinate substituting for
the azimuthal variable:
$$
\eqalignno{&\cos\theta = {z\over \rho} = -{\sigma\over a} & (2.9) \cr
&\rho^2 - \sigma^2 = r^2-a^2 ~. & (2.10) \cr}
$$
\no Later we shall verify that the imaginary displacement, $a$, in (2.7)
measures the specific angular momentum.
In order to establish Eq. (2.5) we must satisfy the simultaneous field
equations
$$
\eqalignno{&R_{\mu\nu} = K(\theta^e_{\mu\nu} + \theta^g_{\mu\nu}) & (2.11) \cr
&\partial_\nu F^{A\mu\nu} = J^{A\mu}~, \qquad A = e,g  & (2.12) \cr}
$$
\no Here $K = {8\pi\over c^2}k$ where $k$ is Newton's constant and where
$$
\eqalign{J^e_\mu &= (e,\vec 0) \delta(\vec x) \cr
J^g_\mu &= (g,\vec 0) \delta(\vec x) ~. \cr} \eqno(2.13)
$$
\no The Kerr-Schild metric has the property that the Lorentzian metric
$(\eta^{\mu\nu})$, as well as $g^{\mu\nu}$, may be used to raise the indices
of $F_{\mu\nu}$ and therefore Eq. (2.12) has the familiar Minkowskian
solution.

The possibility of obtaining Minkowskian solutions here is one example of the
use of Lorentz covariant relations to discuss the Kerr-Newman geometry.  It
was noted by G\"urses and G\"ursey$^8$ that the pseudotensor $\hat\tau^\mu_\nu$,
coupling the gravitational field to itself, vanishes in the Kerr-Schild metric
if the null vector $\ell_\mu$ is also geodesic.  As a consequence, there is
the following linear version of the Gupta equation:
$$
\partial_\alpha\partial_\beta\bigl[\eta^{\alpha\beta}g^{\mu\nu}-
\eta^{\mu\alpha}g^{\nu\beta}-\eta^{\nu\alpha}g^{\mu\beta} +
\eta^{\mu\nu}g^{\alpha\beta}\bigr] = 2K\eta^{\mu\lambda}\theta^\nu_\lambda
\eqno(2.14)
$$
\no where $\theta^\nu_\lambda$ is the energy-momentum tensor of the non-gravitational source.  Here we shall show that $\hat\tau^\mu_\nu$ vanishes
even if $\ell_\mu$ is not geodesic, provided that $\theta^\mu_\nu$ is
traceless.

The solution of (2.12) is
$$
\eqalignno{&F^{(A)} = \partial_\mu A^{(A)}_\nu - \partial_\nu A^{(A)}_\mu~,
 \qquad A = e,g & (2.15) \cr
& A^{(e)}_o = e\alpha & (2.16) \cr
& \vec A^{(e)} = \vec\mu^{(e)} \times\vec\nabla\varphi~, \qquad
\vec\mu^e = (0,0,ea) & (2.17) \cr
& A^{(g)}_o = g\alpha & (2.18) \cr
& \vec A^{(g)} = \vec\mu^{(g)} \times\vec\nabla\varphi~, \qquad
\vec\mu^g = (0,0,ga) & (2.19) \cr}
$$
\no where $(\vec\mu^e,\vec\mu^g)$ are the dipole moments respectively 
associated with the electric and magnetic charges.  Here $\alpha$ and 
$\varphi$ are the same functions for both $A^e_\mu$ and $A^g_\mu$, and$^{12}$
$$
\varphi = {1\over a}\tan^{-1}{\rho\over a}~. \eqno(2.20)
$$
\no At this point both sides of Eq. (2.11) have been expressed in terms of
$\gamma$ as defined in (2.5).  
It remains only to show that the two sides agree.  This step is a
simple extension of the argument in Ref. 7.
\ve

\line{{\bf 3. Horizon and Bound on the Mass.} \hfil}
\s

In order to describe the horizon we transform to polar coordinates
$$
\eqalign{&x+iy = (\rho + ia) e^{i\varphi} \sin\theta \cr
& z = \rho\cos\theta \cr
& \vec\lambda = (\sin\theta\cos\varphi,~\sin\theta\sin\varphi,~\cos
\theta)~. \cr} \eqno(3.1)
$$
\no If $a=0$, $\rho$ is the usual radial coordinate and $\theta$ and
$\varphi$ are the usual polar angles.  In that case $\vec\lambda$ is also a
radial vector but if $a$ does not vanish, the $\vec\lambda$ field defines a
family of curves spiraling into the origin.

Let us also transform to new coordinates $(u,v)$ to eliminate cross-terms in
the Kerr-Schild line element.  Then
$$
ds^2 = E_\rho du^2 - {1\over E_\rho} d\rho^2 - E_\mu dv^2 - {d\mu^2\over
E^2_\mu} \eqno(3.2)
$$
\no where
$$
\eqalign{E_\rho &= {1\over \rho^2}\Delta_\rho \cr
\Delta_\rho &= \rho^2 - 2m\rho + Q^2 \cr
Q^2 &= e^2+g^2+a^2 \cr} \qquad
\eqalign{E_\mu &= {1\over\rho^2} \Delta_\mu \cr
\Delta_\mu &= 1-\mu^2 \cr
\mu~ &= \cos\theta} \eqno(3.3)
$$
\no Here
$$
\eqalign{du &= dt + \bigl[1-(\rho^2+a^2\cos^2\theta)/\Delta]d\rho
-a\sin^2\theta d\varphi \cr
dv &= a~dt-(\rho^2+a^2)d\varphi~. \cr} \eqno(3.4)
$$
\no The horizon of the black hole is determined by
$$
\Delta(\rho) = 0~. \eqno(3.5)
$$
\no Then by (3.2), at the horizon, where the red shift is infinite,
$g_{uu} = 0,~g_{\rho\rho}=\infty$.  If $m^2=Q^2$, the radius of the horizon is
$$
\rho=m=Q~. \eqno(3.6)
$$
\no If $m^2<Q^2$,
$$
\Delta(\rho) = (\rho-m)^2 + Q^2-m^2>0 \eqno(3.7)
$$
\no and there is no horizon.

Therefore the minimum value of the mass for which there is a horizon, or the
maximum value for which there is no horizon, is given by
$$
m^2=Q^2=e^2+g^2+a^2 \eqno(3.8)
$$
\no where all quantities are expressed as lengths.  Then the condition for
the existence of a classical black hole is in general
$$
m^2\geq e^2+g^2+a^2~. \eqno(3.9)
$$
\no If $e=g=a=0$, one sees that there is always a Schwarzchild horizon.

This condition may be compared with the Bogomolny relation
$$
m^2\geq e^2+g^2~. \eqno(3.10)
$$
\no In (3.9) there is, as one would expect, an additional contribution from
the energy of rotation since $a$ is proportional to the angular momentum.

Simple duality is built into the metric (3.2) since electric and magnetic
charges appear only in the combination $e^2+g^2$.  The Reissner-Nordstrom
metric, $g=a=0$, may be obtained by setting $Q=e$ and $u=t,~\rho=r,~
v=-r^2d\varphi$.

One may see that the parameter $m$ appearing in the line element is the
Newtonian mass.  By (2.3), (2.5a)
$$
\eqalign{g_{oo} &= \eta_{oo} -2m\ell^2_o \cr
&= 1-2m\biggl(1-{e^2+g^2\over 2m\rho}\biggr) \alpha~. \cr} \eqno(3.11)
$$
\no Eq. (2.5c) may be inverted
$$
\alpha = {\rho\over \rho^2+\sigma^2}~. \eqno(3.12)
$$
\no Then
$$
g_{oo} = 1-2m\alpha +
\biggl({e^2+g^2\over\rho}\biggr)
\biggl({\rho\over\rho^2+\sigma^2}\biggr)~. \eqno(3.13)
$$
\no Asymptotically
$$
g_{oo} \to 1-{2m\over\rho} + {e^2+g^2\over\rho^2} +\ldots~. \eqno(3.14)
$$
\no The coefficient of ${1\over\rho}$ defines the Newtonian mass.  In general
one may show that the distant field,$^{13}$ with the neglect of self-coupling
of the gravitational field, is
$$
g_{oo} \to 1-{2M\over r} + O\biggl({1\over r^3}\biggr) \eqno(3.15)
$$
\no where
$$
M = \int \theta^{oo} d\vec x~. \eqno(3.16)
$$
\no As we shall see there is no self-coupling of the gravitational field
in the Kerr-Schild frame.  Here $\theta^{oo}$ is the density of energy, the
source of the gravitational field.

By (3.14) and (3.15) one would have
$$
M=m~.
$$
\no For a macroscopic body, such as a star, it is not possible to calculate
$M$ by (3.16); but for a Kerr-Schild dyon, the near field is precisely given
and the integral may be carried out.
\ve

\line{{\bf 4. The Einstein Tensor and the Conformal Current.} \hfil}
\s

The general field equations are
$$
G_{\mu\lambda} = R_{\mu\lambda} - {1\over 2} Rg_{\mu\lambda} =
K\theta_{\mu\lambda}~. \eqno(4.1)
$$
\no In the conformal case
$$
\theta^\mu_{~\mu} = 0~. \eqno(4.2)
$$
\no Then (4.1) becomes
$$
R_{\mu\lambda}=K\theta_{\mu\lambda}~. \eqno(4.3)
$$
\no The Kerr-Schild form of the metric implies
$$
\Gamma^\sigma_{\sigma\mu} = 0 \eqno(4.4)
$$
\no since $\sqrt{-g}=1$.

Then the Ricci tensor simplifies
$$
R_{\mu\lambda} = \partial_\sigma\Gamma^\sigma_{\mu\lambda}
-\Gamma^\alpha_{\mu\beta}\Gamma^\beta_{\lambda\alpha}~. \eqno(4.5)
$$
\no It is useful to set
$$
\Gamma^\sigma_{\mu\lambda} = \mathrel{\mathop{\Gamma^{\sigma}}^{\rm 1~~}}_
{\mu\lambda} + \mathrel{\mathop{\Gamma^{\sigma}}^{\rm 2~~}}_
{\mu\lambda}
$$
\no where $\mathrel{\mathop\Gamma^{\rm 1}}$ and
$\mathrel{\mathop\Gamma^{\rm 2}}$, 
which are first and second order in $m$, are
$$
\eqalignno{\mathrel{\mathop{\Gamma^{\sigma}}^{\rm 1~~}}_{\mu\lambda} 
&= {1\over 2}\eta^{\sigma\tau}
\bigl(\partial_\mu h_{\tau\lambda}+\partial_\lambda h_{\mu\tau}-
\partial_\tau h_{\mu\lambda}) & (4.6a) \cr
\mathrel{\mathop{\Gamma^{\sigma}}^{\rm 2~~}}_{\mu\lambda} 
&= {1\over 2}(2m\ell^\sigma\ell^\tau)
(\partial_\mu h_{\tau\lambda}+\partial_\lambda h_{\mu\tau}-
\partial_\tau h_{\mu\lambda}) & (4.6b) \cr}
$$
\no with
$$
h_{\mu\lambda} = g_{\mu\lambda} - \eta_{\mu\lambda}~. \eqno(4.7)
$$
\no For the dyonic Kerr-Schild solution it may be shown that$^7$
$$
\eqalignno{\partial\ell_\alpha &= -C\ell_\alpha & (4.8) \cr
\partial_\alpha\ell^\alpha &= -D & (4.9) \cr}
$$
\no where $C$ and $D$ are two scalar functions and
$$
\partial = \ell^\mu\partial_\mu~. \eqno(4.10)
$$
\no Then
$$
\mathrel{\mathop{\Gamma^{\sigma}}^{\rm 2~~}}_{\mu\lambda} 
= -4m^2C\ell^\sigma\ell_\mu\ell_\lambda~.
\eqno(4.11)
$$
\no One may now compute
$$
\partial_\sigma\Gamma_{\mu\lambda}^{~\sigma} = {1\over 2}
\bigl[\partial_\mu j_\lambda + \partial_\lambda j_\mu -
\square h_{\mu\lambda}\bigr] + 4m^2(2C^2+DC-\partial C)
\ell_\mu\ell_\lambda \eqno(4.12)
$$
\no where
$$
\eqalignno{j_\lambda &= \partial^\tau h_{\tau\lambda}
& (4.13) \cr
&= 2m(C+D)\ell_\lambda & (4.14) \cr
\noalign{\hbox{and}}
\partial^\tau &= \eta^{\tau\sigma}\partial_\sigma ~. \cr}
$$

One also finds
$$
\Gamma_{\mu\beta}^{~\alpha}\Gamma_{\lambda\alpha}^{~\beta} =
4m^2F\ell_\mu\ell_\lambda \eqno(4.15)
$$
\no where
$$
2F = 3C^2+\partial_\alpha\ell^\beta\partial_\beta\ell^\alpha-
\partial_\alpha\ell^\beta\partial^\alpha\ell_\beta \eqno(4.16)
$$
\no and by (4.5), (4.2) and (4.15)
$$
R_{\mu\lambda} = {1\over 2}\bigl[\partial_\mu j_\lambda+
\partial_\lambda j_\mu + \square h_{\mu\lambda}\bigr] + 4m^2
(2C^2+DC-\partial C-F)\ell_\mu\ell_\lambda~. \eqno(4.17)
$$
\no By (4.2) and (4.3), $R=0$ and
$$
g^{\mu\lambda}(\partial_\mu j_\lambda+\partial_\lambda j_\mu +
\square h_{\mu\lambda}) = 0 \eqno(4.18)
$$
\no or
$$
(\eta^{\mu\lambda}+2m\ell^\mu\ell^\lambda)
(\partial_\mu j_\lambda+\partial_\lambda j_\mu +
\square h_{\mu\lambda}) = 0~. \eqno(4.19)
$$
\no Since terms of the first and second order in $m$ separately vanish, we
have
$$
\eta^{\mu\lambda}(\partial_\mu j_\lambda + \partial_\lambda j_\mu +
\square h_{\mu\lambda}) = 0
$$
\no or
$$
\partial_\mu j^\mu = 0~. \eqno(4.20)
$$
\no Therefore $j^\mu$ is conserved as a consequence of the conformal invariance
of the source.

By (4.14) and (4.20), we have
$$
\partial(C+D) = CD+D^2~. \eqno(4.21)
$$
\no We may also show
$$
\partial(C-D) = CD-\partial_\alpha\ell^\beta\partial_\beta\ell^\alpha~.
\eqno(4.22)
$$

The mixed tensor $R^\mu_{~\lambda}$ is much simpler:
$$
R^\mu_{~\lambda} = g^{\mu\sigma}R_{\sigma\lambda} =
(\eta^{\mu\sigma}+2m\ell^\mu\ell^\sigma)R_{\sigma\lambda}~. \eqno(4.23)
$$
\no One finds
$$
R^\mu_{~\lambda} = (\mathrel{\mathop R^{\rm 1}} +
\mathrel{\mathop R^{\rm 2}})_{~\lambda}^\mu \eqno(4.24)
$$
\no where
$$
\eqalignno{\mathrel{\mathop{R^{\mu}}^{\rm 1~}}_{\lambda} &= {1\over 2}
\bigl[\partial^\mu j_\lambda + \partial_\lambda j^\mu + \eta^{\mu\sigma}
\square h_{\sigma\lambda}\bigr] & (4.25) \cr
\mathrel{\mathop {R^{\mu}}^{\rm 2~}}_{\lambda} 
&= 2m^2\bigl[\partial_\alpha\ell^\beta
\partial_\beta\ell^\alpha-\partial_\alpha\ell^\beta\partial^\alpha\ell_\beta +
3C^2-2F\bigr]\ell^\mu\ell_\lambda & (4.26) \cr}
$$
\no By (4.16)
$$
\mathrel{\mathop {R^{\mu}}^{\rm 2~}}_{\lambda} = 0~. \eqno(4.27)
$$
\no Therefore the mixed Ricci tensor, which is the same as the Einstein
tensor, is simply
$$
R^\mu_{~\lambda} = {1\over 2}
\bigl[\partial^\mu j_\lambda + \partial_\lambda j^\mu -
\eta^{\mu\sigma}\square h_{\sigma\lambda}\bigr]~. \eqno(4.28)
$$
\ve

\line{{\bf 5. The Einstein Pseudo-Energy-Momentum Tensor.} \hfil}
\s

The generally covariant conservation law, namely
$$
G^{\mu\lambda}_{~|\lambda} = K\theta^{\mu\lambda}_{~|\lambda} = 0
\eqno(5.1)
$$
\no implies the conservation equation:
$$
\partial_\mu(\hat\theta^\mu_{~\lambda} + \hat\tau^\mu_{~\lambda}) = 0
\eqno(5.2)
$$
\no where the circumflex indicates the corresponding tensor density
(multiplication by $\sqrt{-g}$).

Here $\hat\theta^\mu_{~\lambda}$ is the energy-momentum tensor that is the
source of the gravitational field and $\hat\tau^\mu_{~\lambda}$ is the
contribution of the gravitational field itself.  Since $\hat\tau^\mu_{\lambda}$
is not a tensor, it may vanish in one frame without vanishing in all frames.

Since $\sqrt{-g}=1$, for the Kerr-Schild metric, the circumflex may be
dropped.

The total (pseudo) energy-momentum tensor, including the contributions of
both the source field and the gravitational field, namely
$$
\Theta^\mu_{~\lambda} = \theta^\mu_{~\lambda} +
\tau^\mu_{~\lambda} \eqno(5.3)
$$
\no may be expressed in the Einstein form
$$
\Theta_{E~\lambda}^{~\mu} = \partial_\gamma h^{\gamma\mu}_{~~\lambda}
\eqno(5.4)
$$
\no where
$$
\eqalignno{h^{\gamma\mu}_{~~\lambda} &= {1\over K} g^{\mu\beta}
{\partial\Gamma\over\partial(\partial_\gamma g^{\lambda\beta})} & (5.5) \cr
\Gamma~~~ &= (\Gamma_\sigma\Gamma_{\alpha\beta}^{~~\sigma} -
 \Gamma^{~~\lambda}_{\alpha\mu}
\Gamma_{\beta\lambda}^{~~\mu}) g^{\alpha\beta} & (5.6) \cr}
$$
\no Using the Kerr-Schild metric one finds
$$
\eqalignno{h^{\gamma\mu}_{~~\lambda} &= {1\over K} g^{\mu\beta}
{\partial(\Gamma_{\varphi\tau}^{~~\sigma}\Gamma_{\psi\sigma}^{~~\tau}
g^{\varphi\psi})\over \partial(\partial_\gamma g^{\lambda\beta})} & (5.7) \cr
&= {1\over K} g^{\mu\beta}\Gamma_{\lambda\beta}^{~~\gamma} & (5.8) \cr}
$$
\no and by (5.4)
$$
\Theta_{E~\lambda}^{~\mu} = {1\over 2K}
(\partial_\lambda j^\mu + \partial^\mu j_\lambda - \eta^{\mu\sigma}
\square h_{\sigma\lambda}) \eqno(5.9)
$$
\no where the covariant and contravariant indices are related by the Lorentz
metric.

By (4.28) and (5.9) one now has
$$
R^\mu_{~\lambda} = K\Theta_{E~\lambda}^{~\mu} =
K(\theta^\mu_{~\lambda} + \tau^\mu_{~\lambda})~. \eqno(5.10)
$$
\no But
$$
R^\mu_{~\lambda} = K\theta^\mu_{~\lambda}~.
$$
\no It follows that
$$
\eqalignno{\Theta_{E~\lambda}^{~\mu} &= \theta^\mu_{~\lambda} & (5.11a) \cr
\noalign{\hbox{or}}
\tau^\mu_{~\lambda}~ &= 0~. & (5.11b) \cr}
$$
\no Hence the gravitational p.e.m.t. vanishes in this metric.

This result depends only on (4.8) and (4.9) and is therefore more general than
the theorem of Ref. 8 which seems to require that the null vector $\ell_\mu$
be geodesic as well, i.e., that $C=0$ in (4.8).  As shown in Ref. 7 (4.8) and
(4.9) hold for the charged (Kerr-Newman) case where $C$ and $D$ satisfy
$$
\eqalignno{ C\ell_o &= {1\over 2} |\gamma|^2 - \alpha\ell_o^2 & (5.12) \cr
D\ell_o &= {1\over 2} |\gamma|^2 + \alpha\ell_o^2~. & (5.13) \cr}
$$
\ve

\line{{\bf 6. The Landau p.e.m.t.} \hfil}
\s

The Landau-Lipshitz prescription for the total p.e.m.t. is
$$
\Theta_L^{\mu\lambda} = {1\over 2K} \partial_\sigma h^{\mu\lambda\sigma}
\eqno(6.1)
$$
\no where
$$
h^{\mu\lambda\sigma} = \partial_\rho\bigl[g^{\mu\lambda} g^{\sigma\rho} -
g^{\mu\sigma} g^{\lambda\rho}\bigr] \eqno(6.2)
$$
\no or
$$
\eqalign{\Theta_L^{\mu\lambda} &=
{1\over 2K} \partial_\sigma \partial_\rho \bigl[ g^{\mu\lambda}
g^{\sigma\rho}-g^{\mu\sigma} g^{\lambda\rho}\bigr] \cr
&= {1\over 2K} \partial_\sigma\partial_\rho
\bigl[\eta^{\mu\lambda} h^{\sigma\rho} + h^{\mu\lambda} \eta^{\sigma\rho} -
\eta^{\mu\sigma} h^{\lambda\rho}-\eta^{\lambda\rho} h^{\mu\sigma}\bigr] \cr
&= -{1\over 2K} \bigl[\partial^\mu j^\lambda + \partial^\lambda j^\mu -
\square h^{\mu\lambda} \bigr]~. \cr} \eqno(6.3)
$$
\no This expression for $\Theta^{\mu\lambda}$ has the desired properties of
symmetry and vanishing Lorentz covariant divergence:
$$
\eqalignno{& \Theta^{\mu\lambda}_L = \Theta^{\lambda\mu}_L & (6.4) \cr
& \partial_\lambda \Theta_L^{\mu\lambda} = 0~. & (6.5) \cr}
$$
\no $\Theta_L^{\mu\lambda}$ is more useful than $\Theta_E^{\mu\lambda}$ since
it permits, by virtue of its symmetry, the easy calculation of a conserved
angular momentum.  By (6.3) the mixed tensor with respect to the Lorentz
metric is
$$
\eqalignno{\mathrel{\mathop\Theta^{\rm o}}_{L~\lambda}^{~\mu} 
&= \eta_{\tau\lambda}
\Theta^{\mu\sigma} & (6.8a) \cr
&= -{1\over 2K}\bigl[\partial^\mu j_\lambda + \partial_\lambda j^\mu
+ \eta^{\mu\sigma} \square h_{\sigma\lambda}\bigr] & (6.8b) \cr}
$$
\no so that the mixed Landau and Einstein tensors agree:
$$
\mathrel{\mathop\Theta^{\rm o}}_{L~\lambda}^{~\mu} 
= \Theta_{E~\lambda}^{~\mu} ~. \eqno(6.9)
$$
\no On the other hand, if the index is lowered by the Kerr-Schild metric,
rather than by the Lorentz metric, one finds
$$
\eqalignno{\Theta_{L~\lambda}^{~\mu} &= g_{\lambda\sigma}
\Theta^{\mu\sigma}_\lambda & (6.10a) \cr
&= \Theta_{E~\lambda}^{~\mu} - (2m)^2 {1\over 2K}
\bigl[C^2-D^2-\ell_\lambda\square\ell^\lambda\bigr]~. & (6.10b) \cr}
$$
\no Although the mixed tensors agree with respect to only the Lorentz
metric, there is a modified Landau p.e.m.t. introduced in Ref. 8, which agrees
as a mixed tensor with the Einstein p.e.m.t. provided that one also uses
the Kerr-Schild metric, namely:
$$
\Theta'^\mu _{L\nu} = -{1\over 2K} \partial_\rho
g_{\nu\sigma}\partial_\lambda(g^{\sigma\mu}g^{\rho\lambda} -
g^{\sigma\rho}g^{\mu\lambda})~. \eqno(6.11)
$$

Then the Einstein and Landau expressions reduce to the same simple form
subject to (4.8), again extending the result of Ref. 9 which requires
$C=0$.

Finally
$$
\partial_\mu\Theta^\mu_{~\nu} = 0
$$
\no implies
$$
\square j_\nu = j_\nu~. \eqno(6.12)
$$
\ve

\line{{\bf 7. Calculation of Mass.} \hfil}
\s

Since the energy density is a perfect divergence, the total energy may be
calculated as the flux through a closed surface at infinity, just as the
electric charge may be found from a similar surface integral.  Since the
closed surface is taken at infinity, the metric may be chosen Lorentzian in
the surface integral.  The metric (2.3) has this property since $\ell_o^2\to 0$.  One commonly takes the closed surface to be spherical.  For our purposes,
however, it is more convenient to take this surface to be $\rho=\hbox{constant}$
instead of $r=\hbox{constant}$.  Then we need the covariant form of Gauss'
theorem:
$$
\int\!\!\int\!\!\int_V F^s_{~|s} dV = \int\!\!\int_S F^s\lambda_s dS \eqno(7.1)
$$
\no where
$$
\eqalignno{F^s_{~|s} &= {1\over\sqrt{g^{(3)}}} \partial_s
\sqrt{g^{(3)}} F^s & (7.2) \cr
dV~~ &= \sqrt{g^{(3)}} dx^1dx^2dx^3 & (7.3) \cr
dS~~ &= \sqrt{g^{(2)}} d\varphi^1 d\varphi^2 & (7.4) \cr}
$$
\no Here $\sqrt{g^{(3)}}=1$ but $\sqrt{g^{(2)}}$ must be computed for an
ellipsoidal surface of constant $\rho$.

By (3.1) we have
$$
\eqalign{x &= (\rho\cos\varphi-a\sin\varphi)\sin\theta \cr
y&= (a\cos\varphi+\rho\sin\rho)\sin\theta \cr
z&=\rho\cos\theta \cr} \eqno(7.5)
$$
\no Then
$$
g^{(2)}_{k\ell} = \sum^3_1 {\partial x^s\over\partial\varphi^k}
{\partial x^s\over\partial\varphi^\ell}~, \qquad k,\ell=1,2 \eqno(7.6)
$$
\no where
$$
\varphi^1=\varphi~, ~~\varphi^2=\theta \eqno(7.7)
$$
\no and
$$
\sqrt{g^{(2)}} = \bigl[(\rho^2+a^2)(\rho^2+\sigma^2)\bigr]^{1/2}
\sin\theta~. \eqno(7.8)
$$
\no By (7.4)
$$
dS = \bigl[(\rho^2+a^2)(\rho^2+\sigma^2)\bigr]^{1/2}
\sin\theta~d\theta d\varphi~. \eqno(7.9)
$$
\no Also
$$
\eqalign{\sigma~&= -a\cos\theta \cr
d\sigma &= a\sin\theta~d\theta~. \cr} \eqno(7.10)
$$
\no Then
$$
dS = {1\over a}\bigl[(\rho^2+a^2)(\rho^2+\sigma^2)\bigr]^{1/2}
d\sigma d\varphi~. \eqno(7.11)
$$

Since$^7$
$$
-\lambda^s\partial_s = {\partial\over\partial\rho}~. \eqno(7.12)
$$
\no $\lambda^s$ is the inward normal to surfaces of constant $\rho$.  Note also
$$
\eqalign{\theta &= 0\to\sigma = -a \cr
\theta &= \pi\to\sigma = a~. \cr} \eqno(7.13)
$$
\no Finally by (7.1)
$$
\int\!\!\int\!\!\int F^s_{~|s} dV = {2\pi\over a}
(\rho^2+a^2)^{1/2} \int^a_{-a} \lambda_sF^s(\rho,\sigma)
(\rho^2+\sigma^2)^{1/2} d\sigma~. \eqno(7.14)
$$
\no This is obviously conserved since all fields are time independent.
In general this expression would be conserved only if the total flux
through the boundary surface vanishes.
\ve

\line{{\bf 8. Landau Mass.} \hfil}
\s

We have for the energy-momentum vector
$$
P^\mu = \int \Theta^{\mu o} dV~. \eqno(8.1)
$$
\no The p.e.m.t. is by (6.1)
$$
\Theta^{\mu\lambda} = {1\over 2K} \partial_\sigma h^{\mu\lambda\sigma} \eqno(8.2)
$$
\no and the mass is
$$
\eqalignno{P^o &= {1\over 2K} \int \partial_k h^{ook} dV & (8.3) \cr
&= {1\over 2K} \int h^{ook} \lambda_k dS~. & (8.4) \cr}
$$

By (6.2)
$$
\eqalignno{ h^{ook} &= \eta^{oo}j^k+\eta^{k\ell}\partial_\ell h^{oo} & (8.5) \cr
&= 2m\bigl[-(C+D)\ell^k+\partial^k\ell_o^2\bigr] & (8.6) \cr
\noalign{\hbox{by (4.14)}}
&= 2m\bigl[-(\alpha^2+\rho^2)\lambda^k+\partial^k\alpha(1-\epsilon/\rho)\bigr]
& (8.7) \cr}
$$
\no by (5.12) and (5.13) and by (2.5a) where
$$
\epsilon = {e^2+g^2\over 2m}~. \eqno(8.8)
$$
\no Then
$$
\lambda_k h^{ook} = 2m\biggl[+\alpha^2+\beta^2+\partial\alpha-
{\epsilon\over\rho} \partial\alpha + {\epsilon\alpha\over\rho^2}
\partial\rho\biggr]
$$
\no where
$$
\partial = \lambda_k\partial^k~.
$$
\no But
$$
\eqalignno{\partial\gamma &= \gamma^2 & (8.9) \cr
\partial\alpha &= \alpha^2-\beta^2 & (8.10) \cr
\partial\rho &= -1 & (8.11) \cr}
$$
\no Then
$$
\lambda_k h^{ook} = 2m\biggl[+2\alpha^2 -
\epsilon\biggl({\alpha\over\rho^2} + {\alpha^2-\beta^2\over\rho}\biggr)
\biggr]~. \eqno(8.12)
$$
\no Here
$$
\eqalign{ {\alpha\over\rho^2} + {\alpha^2-\beta^2\over\rho} &=
{1\over\rho}\biggl({\alpha\over\rho}+\alpha^2-\beta^2\biggr) =
{1\over\rho}\biggl({1\over\rho^2+\sigma^2}+\alpha^2-\beta^2\biggr) \cr
&= {1\over\rho} (\alpha^2+\beta+\alpha^2-\beta^2) \cr
&= {2\alpha^2\over\rho}~. \cr} \eqno(8.13)
$$
\no Then
$$
\eqalign{\lambda_k h^{ook} &= 2m\biggl[+2\alpha^2-{2\alpha^2\over\rho}
\epsilon\biggr] \cr
&= 4m\alpha^2\biggl(1-{\epsilon\over\rho}\biggr)~. \cr} \eqno(8.14)
$$
\no Hence
$$
P^o = {2m\over K}\biggl(1-{\epsilon\over\rho}\biggr)
\int \alpha^2 dS~. \eqno(8.15)
$$
\no This surface integral is
$$
\eqalignno{\int\alpha^2 dS &= {1\over a}\int^{2\pi}_0\int^a_{-a}
\alpha^2\bigl[(\rho^2+a^2)(\rho^2+\sigma^2)\bigr]^{1/2}
d\sigma d\varphi \cr
&= 2\pi(\rho^2+a^2)^{1/2}{1\over a} \int^a_{-a}
{\rho^2\over(\rho^2+\sigma^2)^2}(\rho^2+\sigma^2)^{1/2} d\sigma & (8.16) \cr
&= 4\pi~. & (8.17) \cr}
$$
\no Hence
$$
M(\rho) = {m\over K}\biggl(1-{e^2+g^2\over em\rho}\biggr)~. \eqno(8.18)
$$
\no According to this last equation
$$
\eqalignno{& M(\rho) \leq 0~~~\hbox{if} ~~~ \rho\leq {e^2+g^2\over 2m}
& (8.19) \cr
& M(\infty) = {m\over K}~. & (8.20) \cr}
$$

One may interpret (8.19) and (8.20) by assigning an electromagnetic radius
\break
$(e^2+g^2)/2m$ to this ``particle" since all of the 
positive mass lies outside this
radius.  The limiting relation (8.20) may be interpreted as a statement of the
equivalence principle.

One may be surprised that the angular momentum does not contribute directly
to $M$, but it does determine $M$ indirectly since (8.20) together with (3.8) requires
$$
K^2M^2 = e^2+g^2+a^2 \eqno(8.21)
$$
\no where $M$ is the mass at which the horizon appears.  If the mass $M$ is
greater than $m$, the radius of the horizon is given by
$$
\rho^2_h - 2m\rho_h+Q^2 = 0 \eqno(8.22)
$$
\no or
$$
\rho^2_h -2m\rho_h + 2m\rho_\ell + a^2 = 0 \eqno(8.23)
$$
\no where the electromagnetic radius is
$$
\rho_\ell = (e^2+g^2)/2m~. \eqno(8.24)
$$
\no Hence
$$
2m(\rho_h-\rho_\ell) = \rho_h^2 + a^2>0~. \eqno(8.25)
$$
\no Therefore the electromagnetic radius is always shielded by the
horizon.
\ve

\line{{\bf 9. Angular Momentum.} \hfil}
\s

In terms of the Landau energy momentum tensor the angular momentum is
$$
J^{\alpha\mu} = \int\bigl(x^\alpha\Theta_L^{\mu\nu}-x^\mu
\Theta_L^{\alpha\nu}\bigr) dS_\nu \eqno(9.1)
$$
\no where $dS_\nu$ is an element of a 3-dimensional hypersurface.  By (8.2)
$$
J^{\alpha\mu} = \int\bigl(x^\alpha\partial_\sigma h_L^{\mu\nu\sigma}-
x^\mu\partial_\sigma h_L^{\alpha\nu\sigma}\bigr) dS_\nu \eqno(9.2)
$$
\no where
$$
dS_\nu = {1\over 3!} \epsilon_{\nu\alpha\beta\gamma} dx_1^\alpha
dx_2^\beta dx_3^\gamma~. \eqno(9.3)
$$

If all fields are time-independent then
$$
\eqalignno{J^{ik} &= \int 
\bigl(x^i\partial_s h_L^{kos} - x^k\partial_s h_L^{ios}\bigr) d\vec x 
& (9.4) \cr
&= \int\bigl\{\partial_s(x^ih_L^{kos}-x^kh_L^{ios})-
(h^{koi}-h^{iok})\bigr\} d\vec x \cr
&= I^{ik} + II^{ik}~ & (9.5) \cr}
$$
\no where
$$
I^{ik} = \int (x^ih_L^{kon}-x^kh^{ion}) dS_n \eqno(9.6)
$$
\no and
$$
II^{ik} = \int (h_L^{iok}-h_L^{koi}) d\vec x~. \eqno(9.7)
$$
\no Here $dS_n$ is an element of a 2-dimensional surface.

In (9.6) and (9.7) $h_L^{iok}$ is the Landau tensor:
$$
h^{iks} = {1\over 2K} {\partial\over\partial x^t} H^{ikst} \eqno(9.8)
$$
\no where
$$
H^{ikst} = g^{ik} g^{st}-g^{is}g^{kt}~. \eqno(9.9)
$$
\no Here we have used the Kerr-Schild metric by setting $\sqrt{-g}=1$.
$II^{ik}$ may be transformed to a surface integral by (9.8)
$$
II^{ik} = {1\over 2K} \int (H^{ioks}-H^{kois})\lambda_s dS \eqno(9.10)
$$
\no where the volume in (9.7) is bounded by a surface of constant $\rho$
in (9.10).  Since these surfaces are normal to the $\lambda_s$ vector field,
the integral $I^{ik}$ may be expressed in the following way:
$$
I^{ik} = \int (x^ih_L^{kos}-x^kh_L^{ios})\lambda_s dS~. \eqno(9.11)
$$
\no In (9.10) and (9.11) the element of area on the ellipsoidal $(\rho)$
surface is $dS$.

The integrand of (9.10) is
$$
\lambda_s(H^{ioks}-H^{kois}) = 2m(\ell^o)^2
(\lambda^i\lambda^k-\lambda^k\lambda^i) = 0 \eqno(9.12)
$$
\no where $H^{ioks}$ is reduced by (9.9) and the Kerr-Schild metric.  Then
$$
II^{ik} = 0~. \eqno(9.13)
$$
\no The integral $I^{ik}$ may be evaluated as follows:
$$
\eqalignno{h_L^{kos}~~~ &= {m\over K}\biggl[{\partial\ell^o\over\partial x^t}
(\ell^k\eta^{st}-\ell^t\eta^{sk})+\ell^o(\partial^s\ell^k-\eta^{sk}
\partial_t\ell^t)\biggr] \cr
h_L^{kos}\lambda_s &= {m\over K}
\biggl[\ell^o{\partial\ell^o\over\partial x^t}(\lambda^k\lambda^t-\lambda^t
\lambda^k) + \ell^o(\lambda_s\partial^s\ell^k-\lambda^k\partial_t\ell^t)\biggr]
& (9.14) \cr
&= {m\over K}[\ell^o(\lambda_s\partial^s\ell^k-\lambda^k\partial_t\ell^t)] \cr
&= {m\over K} (D-C)\ell^k & (9.15) \cr}
$$
\no and
$$
(x^ih_L^{kos}-x^kh_L^{ios})\lambda_s =
{m\over K} \ell^o(D-C)(x^s\lambda^k-x^k\lambda^i)~.
$$
\no Then
$$
I^{ik} = {m\over K}\int \ell_o(D-C)(x^i\lambda^k-x^k\lambda^i)dS~.
\eqno(9.16)
$$

Since the imaginary displacement is along $z$ we consider $I^{12}$ and 
compute$^7$
$$
x^1\lambda^2-x^2\lambda^1 = -a\biggl(1-{\sigma^2\over a^2}\biggr)~.
\eqno(9.17)
$$
\no We also need
$$
\eqalign{&\ell^o(D-C) = 2\alpha\ell_o^2 \cr
&\ell_o^2 = \alpha(1-\epsilon/\rho)~. \cr} \eqno(9.18)
$$
\no by (5.12) and (2.5a).  Then
$$
I^{12} = -{2m\over K} a\biggl(1-{\epsilon\over\rho}\biggr)
\biggl\{\int\alpha^2 dS-{\rho^2\over a^2}\int \beta^2 dS\biggr\}~. \eqno(9.19)
$$
\no The first integral is known from (8.17).  The second integral is
$$
\eqalignno{\int\beta^2 dS &= {(\rho^2+a^2)^{1/2}\over a} \int^{2\pi}_o
\int^a_{-a} \beta^2(\rho^2+\sigma^2)^{1/2} d\sigma d\varphi & (9.20) \cr
&= {2\pi\over a} (\rho^2+a^2)^{1/2} \int^a_{-a}
{\sigma^2\over (\rho^2+\sigma^2)^{3/2}} d\sigma & (9.21) \cr
&= -4\pi + {2\pi\over a} (\rho^2+a^2)^{1/2}
\ln {a+(\rho^2+a^2)^{1/2}\over -a+(\rho^2+a^2)^{1/2}}~. & (9.22) \cr}
$$
\no By (8.17) and (9.22) the total angular momentum is
$$
J_3 = I^{12} = -{4\pi m\over K}
\biggl(1-{\epsilon\over\rho}\biggr)(\rho^2+a^2)^{1/2}
\biggl\{{2\over a}(\rho^2+a^2)^{1/2}-{\rho^2\over a^2} \ln
{a+(\rho^2+a^2)^{1/2}\over -a+(\rho^2+a^2)^{1/2}}\biggr\}~. \eqno(9.23)
$$
\no Again
$$
\eqalign{J_3(\rho) &\leq 0 ~~~\hbox{if} ~~~ \rho \leq {\ell^2 + g^2\over 2m} \cr
J_3(\infty) &= -{2\over 3}{ma\over k}~. \cr} \eqno(9.24)
$$
\no We may regard the angular momentum as confined to the space outside of
the ``electromagnetic radius".

We finally have
$$
J_3/M = {2a\over 3}~. \eqno(9.25)
$$
\no Similarly we find
$$
\eqalign{x^1\lambda^3-x^3\lambda^1 &= \lambda_2\lambda_3 \cr
x^2\lambda^3-x^3\lambda^2 &= -\lambda_1\lambda_3 \cr} \eqno(9.26)
$$
\no Utilizing (9.16) and (3.1) one may show that $J_1$ and $J_2$ vanish.
\ve

\line{{\bf 10. Solitons.} \hfil}

The particle-like solutions so far discussed in this paper, as well as the
string-derived solution, being descendants of the Schwarzschild solutions,
all exhibit  central singularities.  Since these structures are also all
time independent, the theorem of Penrose, Hawking, and Geroch does not directly
apply.  In any case this theorem requires certain conditions on the
energy-momentum tensor that do not seem to be required by any fundamental
principle.$^{14}$  There is thus apparently no necessary requirement of a
central singularity and there are certainly macroscopic examples in which
the gravitational attraction is compensated in steady state structures without
central singularities.

It is known that singularity free solitons may be constructed at the special
relativistic level.$^{15,16}$  The fields which are codetermined in these
known structures remain finite with flat tangents at the origin and in general
exhibit nodal behavior before vanishing at large distances.  In this respect
the constituent fields resemble the wave function of atomic and nuclear
physics.

In looking for a replication of these or similar structures at the general
relativistic level, two examples naturally come to mind and illustrate the
complexity of the new situation.  The first of these is formed by
coupling the gravitational field to a gauge structure such as the 
Prasad-Somerfield soliton.  The coupling is formally accomplished by 
replacing $\partial_\mu$ by $\nabla_\mu = \partial_\mu+\Gamma_\mu$ 
in the special
relativistic equations.  Since the  
Prasad-Somerfield solution itself already 
contains $1/r$ singularities, however, it is unlikely that the new
soliton is singularity free.

As a second example we consider the simplest possibility, namely the
gravitational field coupled to a non-linear scalar field.  It is known that
the nonlinear scalar field may be used to construct a singularity free
soliton at the special relativistic level.$^{15,16}$  We must, however, now satisfy the gravitational
field equations as well.

Let the Lagrangian of the scalar field be
$$
\eqalign{L &= T-V \cr
T &= {1\over 2} g^{\mu\lambda} \partial_\mu\psi\partial_\lambda\psi \cr
V &= f(\psi)~. \cr} \eqno(10.1)
$$
\no Then the energy-momentum tensor is
$$
\theta_{\mu\lambda} = {\partial L\over \partial g^{\mu\lambda}}-
{1\over 2} g_{\mu\lambda} L~. \eqno(10.2)
$$
\no The equation of motion of the $\psi$ field is
$$
g^{\mu\lambda} \nabla_\mu\partial_\lambda\psi +
{\partial f(\psi)\over \partial\psi} = 0~. \eqno(10.3)
$$

Since the conformal assumption may already imply a central singularity we
do not make this assumption and therefore adopt the following gravitational
field equations
$$
R_{\mu\lambda} = K\Theta_{\mu\lambda} \eqno(10.4)
$$
\no where
$$
\Theta_{\mu\lambda} = \theta_{\mu\lambda} - {1\over 2} \theta g_{\mu\lambda}
\eqno(10.5)
$$
\no and
$$
\theta \not= 0~. \eqno(10.6)
$$
\no Assume spherical symmetry and let $\lambda_k$ be the unit radial vector.
As the simplest ansatz let us again assume a Kerr-Schild metric.
Then
$$
\eqalign{\Theta_{oo} &= m\varphi f(\psi) - {1\over 2} f(\psi) \cr
\Theta_{ok} &= m\varphi f(\psi) \lambda_k \cr
\Theta_{jk} &= {1\over 2} \delta_{jk} f(\psi) + {1\over 2}
\lambda_j\lambda_k
\biggl[\biggl({d\psi\over dr}\biggr)^2 + 2m\varphi f(\psi)\biggr]~. \cr}
\eqno(10.7)
$$
\no and 
$$
\eqalign{R_{oo} &= -m\nabla^2\varphi + 2m^2\varphi\nabla^2\varphi \cr
R_{ok} &= 2m^2(\varphi\nabla^2\varphi) \lambda_k \cr
R_{jk} &= \delta_{jk} 2m\biggl({1\over r}{d\varphi\over dr} +
{\varphi\over r^2}\biggr) + \lambda_j\lambda_k
\biggl[m\biggr({d^2\varphi\over dr^2}-{2\varphi\over r^2}\biggr)
+2m^2\varphi\nabla^2\varphi\biggr] \cr} \eqno(10.8)
$$
\no where
$$
\varphi = \ell_o^2~. \eqno(10.9)
$$
\no The gravitational equations of motion are now
$$
\eqalignno{-&m\nabla^2\varphi+2m^2\varphi\nabla^2\varphi =
K[m\varphi f(\psi) - {1\over 2} f(\psi)]~~~~~~~~~~~~~~~~~~~(oo) & (10.10) \cr
&2m^2\varphi\nabla^2\varphi = Km\varphi f(\psi) ~~~~~~~~~~~~~~~~~~~~~~~~~~~~~~~~
~~~~~~~~~~~~~
(ok) 
& (10.11) \cr
& 2m\biggl
({1\over r}{d\varphi\over dr}+{\varphi\over r^2}\biggr) =
{K\over 2} f(\psi) ~~~~~~~~~~~~~~~~~~~~~~~~~~~~~~~~~~~~~~~~
(kk) & (10.12) \cr
&m\biggl({d^2\varphi\over dr^2}-{2\varphi\over r^2}\biggr) +
2m^2\varphi\nabla^2\varphi = {K\over 2}
\biggl[\biggl({d\psi\over dr}\biggr)^2 + 2m\varphi f(\psi)\biggr]~~(jk) &
(10.13) \cr}
$$
\no These equations must be satisfied simultaneously with (10.3) subject to
the solitonic boundary conditions requiring that $\varphi(r)$ and $\psi(r)$
vanish at infinity and remain finite with flat tangents at $r=0$.  Although
the Kerr-Schild form is versatile enough to be compatible with the energy
momentum tensor of a dyonic field, there is no solitonic $\varphi(r)$ which
is compatible with a solitonic solution $\psi(r)$ of (10.3) for any choice of the free function $f(\psi)$.  This may be shown as follows.

The gravitational equations may be combined in the following way
$$
\eqalignno{(10.10)+(10.11) &\to \nabla^2\varphi = {K\over 2m} f(\psi) &
(10.14) \cr
(10.12)+(10.11) &\to {d^2\varphi\over dr^2} = {2\varphi\over r^2} &
(10.15) \cr
(10.13)+(10.14) &\to {d^2\varphi\over dr^2}-{2\varphi\over r^2}=
{K\over 2m}\biggl({d\psi\over dr}\biggr)^2 & (10.16) \cr}
$$
\no The solution of (10.15) which satisfies boundary
conditions at $r=0$ is $\varphi=ar^2$ but it blows up at $\infty$.  
The second solution $\varphi={a\over r}$ blows up at the origin.  Finally,
$$
(10.15)+(10.16) \to {d\psi\over dr} = 0~. \eqno(10.17)
$$

Equation (10.16) is obviously inconsistent with a solitonic solution of
(10.3).

The failure of these simple choices is not surprising since the source field
and the gravitational field are not parts of a larger structure in these
examples.  In a more promising approach one first introduces a unitary field
generated by a postulated symmetry group.  Then one part of the unitary field
is recognized as the Einsteinian gravitational field while the remaining
part is identified as the matter field.  Such a split is made in the theory of
the superstring and supergravity.

Let us therefore study, as a third example, a total field resulting from the
reduction to four dimensions of a particular superstring theory.  This field
may be described by the following $N=4$ supergravity action.$^{17}$
$$
S=\int d^4x\sqrt{-g}\biggl\{R+{1\over 2} g^{\mu\lambda} \partial_{(\mu}
\psi\partial_{\lambda)}\psi^*-f(\psi)\bigl[F_{\mu\lambda}F^{\mu\lambda}+
G_{\mu\lambda}G^{\mu\lambda}\bigr]\biggr\}~. \eqno(10.18)
$$
\no Here $\psi$ is a complex scalar, whose real part is the dilaton and whose
imaginary part is the axion.  This action is a slight generalization of the
$SU(4)$ version of $N=4$ supergravity where
$$
f(\psi)=e^{-2\psi}~. \eqno(10.19)
$$

Solutions to the field equations arising from (10.18) and (10.19) have been
found by Gibbons and others.  These carry central singularities hidden by a
horizon.  Here we would like to show that again there are no singularity
free solutions for a wide class of gravitational fields.

For simplicity ignore the field associated with the magnetic charge and
therefore set
$$
G_{\mu\lambda}=0~. \eqno(10.20)
$$
\no We adopt the Lagrangian $R+L$ where
$$
L = {1\over 2} g^{\mu\lambda} \nabla_{(\mu}\psi\nabla_{\lambda)}\psi^*-
f_1(\psi)F_{\mu\lambda}F^{\mu\lambda}-f_2(\psi)~.
\eqno(10.21)
$$
\no Here
$$
\nabla_\mu = \partial_\mu-ieA_\mu-\Gamma_\mu \eqno(10.22)
$$
\no and we have also added a second nonlinear term, which may include a mass
term for the scalar field.  Here $\Gamma_\mu$ represents the 
gravitational coupling

One now has the gravitational equation (10.4) and
$$
\Theta_{\mu\lambda} = {1\over 2}\nabla_{(\mu}\psi\nabla^*_{\lambda)}\psi^*-
2f_1(\psi)\theta^{e\ell}_{\mu\lambda}-{1\over 2}f_2(\psi)g_{\mu\lambda}
\eqno(10.23)
$$
\no where
$$
\theta^{e\ell}_{\mu\lambda}=F_{\mu\sigma}(\psi)F_{\lambda}^\sigma(\psi)-
{1\over 4}F_{\alpha\beta}(\psi) F^{\alpha\beta}(\psi)g_{\mu\lambda}~.
\eqno(10.24)
$$
\no The source of $F_{\mu\lambda}(\psi)$ is the charged scalar, $\psi$:
$$
\eqalignno{F_{\mu\lambda} &= \partial_\mu A_\lambda - \partial_\lambda
A_\mu & (10.25) \cr
\square A_\mu &= j_\mu(\psi) & (10.26) \cr}
$$
\no where
$$
j_\mu(\psi) \sim \psi^*\nabla_\mu\psi - \psi\nabla_\mu\psi^*~. \eqno(10.27)
$$
\no The nonlinear equation of motion of the scalar field is
$$
g^{\mu\lambda}\nabla_\mu\nabla_\lambda\psi-
{\partial f_1(\psi)\over\partial\psi} F_{\mu\lambda}(\psi) F^{\mu\lambda}(\psi)
-{\partial f_2(\psi)\over\partial\psi} = 0~. \eqno(10.28)
$$
\no If $\psi$ is complex (10.26) and (10.28) are strongly coupled.  In
addition to (10.26) and (10.28) one has the gravitational equtions
(10.4).

In the paper of Kallosh {\it et al.}, where $\psi$ is real, the ansatz for
$ds$ is
$$
ds^2 = e^{2u}dt^2-e^{-2u} dr^2-R^2 d\Omega \eqno(10.29)
$$
\no and one finds
$$
\eqalignno{e^{2u} &= {(r-r_-)(r-r_+)\over R^2} & (10.30) \cr
e^{2\psi} &= e^{2\psi_o} {r+\Sigma \over r-\Sigma} & (10.31) \cr} 
$$
\no where
$$
R^2 = r^2-\Sigma^2~. \eqno(10.32)
$$
\no The curvature singularity is at $r=|\Sigma|$ which is shielded by the
horizon.  $\Sigma$ is determined by the mass, charges, and asymptotic value
of the dilaton fifeld.

We shall here adopt the Kerr-Schild metric (2.3).  This metric is chosen
because it is able to accommodate the charged rotating source.  Although
it also displays a central singularity when the source of mass and
charge is confined to a point, it is at least {\it a prioiri} possible that
the singularity will disappear if the source of mass and charge is spread out
as it would be if the charged scalar field is also spread out.  We shall
investigate this point by examining the gravitational field equations (10.4).

Let us consider the spherically symmetric non-rotating case (corresponding 
to \break
Reissner-Nordstrom rather than Kerr-Newman).  Then the vector potential
vanishes.

Assume harmonic time-dependence of $\psi$:
$$
\psi = Re^{i\omega t}~. \eqno(10.33)
$$
\no Then
$$
\nabla_o\psi = i(\omega-eA_o)\psi~. \eqno(10.34)
$$
We may rewrite (10.23)
$$
\Theta_{\mu\lambda} = {1\over 2} \nabla_{(\mu}\psi\nabla^*_{\lambda)}\psi^*-
\bigl[a(\psi)\eta_{\mu\lambda}+b(\psi)\ell_\mu \ell_\lambda\bigr] \eqno(10.35)
$$
\no where $a(\psi)$ and $b(\psi)$ are new scalars determined by $f_1(\psi)$,
$f_2(\psi)$, and $\theta^{e\ell}_{\mu\lambda}$.  Then, if $R$ is real,
$$
\Theta_{oo} = (\omega-eA_o)^2R^2-[a(\psi)+b(\psi)\varphi]
\eqno(10.36)
$$
\no where
$$
\varphi=\ell_o^2 \eqno(10.37)
$$
\no and
$$
\eqalign{\Theta_{ok} &= -b(\psi)\varphi\lambda_k \cr
\Theta_{jk} &= ~a(\psi)\delta_{jk}-b(\psi)\varphi\lambda_j\lambda_k+
\biggl({dR\over dr}\biggr)^2 \lambda_j\lambda_k~. \cr} \eqno(10.38)
$$
\no Here
$$
\ell_k = \ell_o\lambda_k \eqno(10.39)
$$
\no and $\lambda_k$ is a unit radial vector.

The Ricci tensor is again given by (10.8).  Then the gravitational equations
become
$$
\eqalignno{-&m\nabla^2\varphi + 2m^2\varphi\nabla^2 \varphi =
K[\tilde\omega^2 R^2-\bigl(a(\psi)+b(\psi)\varphi\bigr)] & (10.40) \cr
&(2m^2\varphi\nabla^2\varphi)\lambda_k = -Kb(\psi)\varphi\lambda_k &
(10.41) \cr
&2m\biggl({\varphi^\prime\over r}+{\varphi\over r^2}\biggr) = Ka(\psi) & 
(10.42) \cr
&m\biggl(\varphi^{\prime\prime}-{2\varphi\over r^2}\biggr)+ 2m^2\varphi\nabla^2\varphi
= K[(R^\prime)^2-b(\psi)\varphi] & (10.43) \cr}
$$
\no where
$$
\tilde\omega = \omega-e A_o~. \eqno(10.44)
$$
\no By (10.41)
$$
\nabla^2\varphi = -{K\over 2m^2} b(\psi)~. \eqno(10.45)
$$
\no By (10.40) and (10.41)
$$
\nabla^2\varphi = -{K\over m} [\tilde\omega^2R^2-a(\psi)]~. \eqno(10.46)
$$
\no By (10.46) and (10.42)
$$
2m\biggl({\varphi^\prime\over r}+{\varphi\over r^2}\biggr) = m\nabla^2\varphi +
K\tilde\omega^2R^2 \eqno(10.47)
$$
\no or
$$
m\biggl[{2\varphi\over r^2}-\varphi^{\prime\prime}\biggr] = K\tilde\omega^2R^2~.
\eqno(10.48)
$$
\no By (10.43)
$$
2m^2\varphi\nabla^2\varphi+Kb(\psi)\varphi =
K\bigl[(R^\prime)^2+\tilde\omega^2R^2\bigr]~. \eqno(10.49)
$$
\no By (10.45)
$$
(R^\prime)^2 + \tilde\omega^2R^2 = 0~. \eqno(10.50)
$$
\no If $R$ is not real, the argument is unchanged but $R$ is replaced by
$|R|$.  Since $|R|$ and $\tilde\omega$ are real,
$$
|R|^\prime = |R| = 0~.  \eqno(10.51)
$$
\no In additiion, by (10.48)
$$
\varphi^{\prime\prime} = {2\varphi\over r^2} \eqno(10.52)
$$
\no with the independent solutions
$$
\varphi \sim r^2 \eqno(10.53)
$$
\no and
$$
\varphi \sim {1\over r}~. \eqno(10.54)
$$
\no $\varphi$ satisfies solitonic boundary conditions at the origin
according to (10.53) and at infinity according to (10.54).  Finally
(10.51) is inconsistent with a non-trivial solution of (10.28).  Once again
it is not possible to find a genuine soliton.

If there actually is a physical basis for associating elementary particles
with singularity free solitons, however, it should not be easy to construct
these structures. 
It would be more reasonable 
to expect success only with fundamenttnal theories having established
physical content.

\ve
\baselineskip=12pt

\line{{\bf References.} \hfil}
\s

\item{1.} G. 't Hooft, Nucl. Phys. B{\bf 79} 276 (1974);
A. M. Polyakov, JETP Lett. {\bf 20}, 194 (1974).
\vskip.3cm

\item{2.} M. K. Prasad and C. M. Sommerfield, Phys. Rev. D{\bf 35}, 780
(1975); E. B. Bogolmolyni, Sci. J. Nucl. Phys. {\bf 24}, 447 (1979).
\vskip.3cm

\item{3.} G. W. Gibbons and C. N. Hull, Phys. Lett. {\bf 109B}, 190 (19820).
\vskip.3cm

\item{4.} J. Harvey and A. Strominger
, hep-th/9504047;
\item{} A. Tseytlin, hep-th/9601177;
\item{} M. Cvetic and A. Tseytlin, hep-th/9512031;
\item{} A. Sen, hep-th/9411187; hep-th 9504147; hep-th/9210050;
\item{} M. Cvetic and D. Youm, htp-th/9512127; hep-th/9507090;
\item{} J. C. Breckenridge, R. C. Myers, A. W. Peet, C. Vafa,
HUTP-96/A005, McGill/96-07, PUPT-1592;
\item{} D. Jatkar, S. Mukherji and S. Panda, hep-th/9601118.
\vskip.3cm

\item{5.} J. Schwinger, Science {\bf 165} 757 (1969).
\vskip.3cm

\item{6.} G. C. Debney, R. P. Kerr, and A. Schild, J. Math. Phys. {\bf 10},
1842 (1969).
\vskip.3cm

\item{7.} R. Finkelstein, J. Math. Phys. {\bf 16}, 1271 (1975);
\item{} S. Einstein and R. Finkelstein, Phys. Rev. D{\bf 15}, 2721 (1977).
\vskip.3cm

\item{8.} M. G\"urses and F. G\"ursey, J. Math. Phys. {\bf 16}, 2385 (1975).
\vskip.3cm

\item{9.} M. M. Schiffer, R. J. Adler, J. Mark, and C. Sheffield,
J. Math. Phys. {\bf 14}, 52 (1973).
\vskip.3cm

\item{10.} E. T. Newman and A. I. Janis, J. Math. Phys. {\bf 6}, 915 (1965);
\item{} E. T. Newman, E. Crouch, K. Chinnapared, A. Exton, A. Prakash, and
K. Torrence, J. Math. Phys. {\bf 6}, 918 (1965).
\vskip.3cm

\item{11.} A. Einstein, Ann. Phys. {\bf 49}, 769 (1916);
\item{} L. D. Landau and E. M. Lifshitz, {\it The Classical Theory of
Fields}, Addison-Wesley, Reading, Mass. (1971), p. 306;
\item{} A. Papapetrov, Proc. Roy. Irish Acad. {\bf A52}, 11 (1948);
\item{} S. N. Gupta, Phys. Rev. {\bf 96}, 1683 (1954);
\item{} C. M\o ller, Ann. Phys. {\bf 4}, 347 (1958);
\item{} J. N. Goldberg, Phys. Rev. {\bf 111}, 315 (1958);
\item{} P. A. M. Dirac, Phys. Rev. Lett. {\bf 2}, 368 (1959).
\vskip.3cm

\item{12.} R. Arnowitt, S. Deser, and C. W. Misner, Phys. Rev. {\bf 122},
947 (1961).
\vskip.3cm

\item{13.} C. Misner, K. Thorne, and J. Wheeler, {\it Gravitation}, 
Freeman, San Francisco, CA (1970).
\vskip.3cm

\item{14.} J. D. Bekenstein, Phys. Rev. D{\bf 11}, 2072 (1975).
\vskip.3cm

\item{15.} R. Finkelstein, R. LeLevier, and M. Ruderman, Phys. Rev. {\bf 83},
326 (1951).
\vskip.3cm

\item{16.} R. Friedberg, T. D. Lee, A. Sirlin, Phys. Rev. D{\bf 13}, 2739
(1976)

\item{17.} R. Kallosh, A. Linde, T. Ortin, A. Peet, and
A. Van Proeyen, SU-ITP-92-13..

\ve

\end

\bye